\documentclass[prd,aps,superscriptaddress,floatfix,nofootinbib,eqsecnum,onecolumn]{revtex4-1}
\pdfoutput=1

\usepackage{amsmath}
\usepackage{amssymb}
\usepackage{physics}
\usepackage{bbold}
\usepackage{graphicx}
\usepackage{verbatim}
\usepackage{hyperref}
\usepackage{empheq}

\usepackage{indentfirst}

\begin{document}

\title{Constraints on Metastable Dark Energy Decaying into Dark Matter}

\author{
J.~S.~T.~de~Souza}
\email{jonathas\_sts@id.uff.br}
\affiliation{Instituto de F\'{\i}sica, Universidade Federal Fluminense,
24210-346 Niter\'oi, RJ, Brazil}

\author{G.~S.~Vicente}\email{gustavo@fat.uerj.br}
\affiliation{Faculdade de Tecnologia, Universidade do Estado do Rio de Janeiro, 27537-000 Resende, RJ, Brazil.}

\author{L.~L.~Graef}
\email{leilagraef@id.uff.br}
\affiliation{Instituto de F\'{\i}sica, Universidade Federal Fluminense,
24210-346 Niter\'oi, RJ, Brazil}

%\author{E.~G.~M.~ Ferreira ??}
%\email{...}
%\affiliation{...}

\begin{abstract} 
We revisit the proposal that an energy transfer from dark energy into dark matter can be described in field theory by a first order phase transition. We analyze a metastable dark energy model proposed in the literature, using updated constraints on the decay time of a metastable dark energy from recent data. The results of our analysis show no prospects for potentially observable signals that could distinguish this scenario from the $\Lambda\textrm{CDM}$. We analyze, for the first time, the process of bubble nucleation in this model, showing  that such model would not drive a complete transition to a dark matter dominated phase even in a distant future. Nevertheless, the model is not excluded by the latest data and we confirm that the mass of the dark matter particle that would result from such a process corresponds to the mass of an axion-like particle, which is currently one of the best motivated dark matter candidates. We argue that extensions to this model, possibly with additional couplings,  still deserve further attention as it could provide an interesting and viable description for an interacting dark sector scenario based in a single scalar field.
\end{abstract}

\maketitle

\tableofcontents

\section{Introduction}%
\label{sec1}

One of the main %MDPI: Please move references 1 and 2 citation in abstract part to main text.
 goals of physical cosmology has been to understand
the nature of the constituents of our Universe, especially dark matter (DM) and dark energy (DE) which, together, constitute nearly 96$\%$ of the total density of the Universe.
Many theoretical models have been developed to explain the nature of the dark Universe~\cite{Sahni:2004ai}.  The~DM component,  due to its clustering properties, has been investigated in the context of several astrophysical and cosmological experiments. Moreover, unlike the case of  DE, it has also been studied at particle physics level, being sought in direct detectors on Earth. In~this context, one of the most interesting possibilities is that the DM can be an axion-like particle (a good review on axion-like dark matter can be found in%MDPI: We removed ``Ref. and Refs.'', Response: I confirm.
~\cite{Ferreira:2020fam}, for~instance), which is one of the main candidates for this component today \footnote{For constraints on DM mass in related contexts see for instance~\cite{Amin:2022nlh, Nadler:2021dft, Irsic:2017yje, Dalal:2022rmp, Powell:2023jns, Semertzidis:2021rxs, Nakai:2022dni, Nakatsuka:2022gaf, QUAX:2020uxy}.}. On~the other hand, the~nature of the DE component remains yet very obscure, although~there have been important advances in modeling its behavior beyond the simple cosmological constant~scenario.

 The explanation for a late-time accelerated phase in the Universe remains a topic of much debate, which is related to the well-known cosmological constant problem~\cite{d28dd6a75fda4c2eb0aa9e85b7da702e, Martin:2012bt}.  In~a different framework,  there are many challenges when trying to embed such models in a more fundamental quantum gravity proposal. As~an example,  we can  mention the 
 Swampland conjecture, which describes  a whole  inhabitable landscape of field theories that are inconsistent with string theory, including the stable de Sitter vacua~\cite{Palti:2019pca, Heisenberg:2018rdu, Heisenberg:2018yae}. In~the process of trying to understand the DE properties, we are faced with the recurring discussion in the literature concerning the issue of the stability of a de Sitter phase. In~the context of the late time Universe, a~stable de Sitter phase has shown either to be hard to achieve from fundamental physics or even to be inconsistent in different theoretical contexts. For~example, it has been subject of a long debate whether a  de Sitter space is unstable due to infrared (IR) effects, as~conjectured in~\cite{Polyakov:2007mm,Polyakov:2012uc, Valiviita:2008iv, Mazur:1986et, Mottola:1985ee}, for~instance. An~instability of a de Sitter phase has also been obtained as a consequence of the backreaction effects of super-Hubble modes. As~shown in several works~\cite{Brandenberger:2018fdd, Abramo:1997hu, Mukhanov:1996ak, Finelli:2001bn, Finelli:2003bp, Marozzi:2006ky, Brandenberger:1999su}, the~backreaction of super-Hubble modes could give a negative contribution to the effective cosmological constant, causing the latter to~relax.

From the observational point of view, one motivation for investigating possibilities beyond the cosmological constant solution is the current tension in $H_{0}$ measurements, which has shown to be alleviated in some quintessence models, as~well as in models with dark sector interaction~\cite{deSa:2022hsh, DiValentino:2019exe, DiValentino:2019ffd, Zhao:2017cud}. Concerning the latter, in~the
framework of field theory, it can be argued that it is rather natural to consider an
interaction between DM and DE, given that they are
fundamental fields of the theory. Interacting models can also be claimed as a proposal for alleviating the coincidence problem~\cite{Weinberg:2000yb}, while also being able to provide a good fit to current data~\cite{Ferreira:2014jhn, Amendola:1999er, Valiviita:2008iv, Abdalla:2009mt, Faraoni:2014vra, He:2008tn, Costa:2013sva, Benetti:2021div}.

In a related framework, when going beyond the cosmological constant scenario, it is natural to investigate the possibility of a metastable DE. As~pointed out  in%MDPI: We removed ``Ref. and Refs.'', please confirm.
~\cite{Li:2019san}, the~remarkable qualitative similarity between the  properties
of the present DE and the  component that supposedly drove inflation
in the very early Universe makes it rather natural to put forward the hypothesis that the current DE can also be metastable~\cite{Li:2019san, Urbanowski:2021waa, Urbanowski:2022iug, Landim:2016isc, Landim:2017lyq,  Stojkovic:2007dw, Greenwood:2008qp, Abdalla:2012ug, Casey:2024jep, Freese:2023fcr}.
In this context, in~\cite{Shafieloo:2016bpk}, and~more recently in~\cite{Li:2019san},  metastable DE  phenomenological models were analyzed, in~which the DE decay rate does not depend on external parameters, being 
assumed to
be a constant depending only on DE intrinsic properties. In~the latter analysis, they considered data from  Pantheon compilation~\cite{Pan-STARRS1:2017jku} in combination
with BAO data from 6dFGS~\cite{Beutler:2011hx}, MGS~\cite{Ross:2014qpa}, BOSS DR12~\cite{BOSS:2016wmc}, eBOSS DR14~\cite{eBOSS:2018yfg}, Ly$\alpha$ \cite{BOSS:2017uab}, and CMB~\cite{Chen:2018dbv,Planck:2018vyg}, and~found that the typical decay time in these scenarios must be many times larger than the age of the~Universe.

One possible theoretical model that can provide a field theory description for the class of phenomenological scenarios considered in the analysis of~\cite{Li:2019san} is the model proposed in~\cite{Abdalla:2012ug}, hereafter referred to as the MDE (Metastable Dark Energy) model.
 In this MDE model, a~positive ``cosmological constant"  is modeled by a nonzero scalar
vacuum energy with a potential of the expected order $V \approx 10^{-47} {\rm GeV}^4$ in the false vacuum. The~potential of the scalar field in this model has a doubly degenerate energy minimum with
small symmetry breaking terms that provide such a small energy difference \footnote{There are examples in which this configuration can naturally appear as, for~example, in~the case of the Wess--Zumino model~\cite{osti_4338791}, which has a double
degenerate bosonic vacuum due to super-symmetry, presumably broken only non perturbatively.}. In~this scenario, the field at the false vacuum represents DE.
 After the field passes the potential barrier, decaying from the false to the true vacuum, its equation of state is no longer that of dark
energy, as~it acquires non-negligible kinetic energy \footnote{Other interesting models which consider a unified dark sector through a single field can be found for instance in~\cite{Brandenberger:2019jfh, Brandenberger:2019pju, Bertacca:2010ct, Frion:2023} and references therein.}. Analogously to what 
happens in the old inflationary scenario, the~transition to
the true vacuum occurs through the formation of bubbles
of a new vacuum~\cite{Callan:1977pt, Coleman:1977py}. Through this process, 
 the energy released in the conversion of the
false vacuum into the true can produce a new component, which has the properties of DM. In the MDE %MDPI: We removed bold, please confirm. Response: I confirm.
 model, there is a single scalar field describing the dark sector of the Universe. Such a field is not responsible for inflation at early times and it has no significant coupling to the standard model sector (or to the inflation), except~through the gravitational interaction. In~addition  to the absence of significant coupling to the standard model particles, despite there being coupling with gravity, we are in the small field regime.  In~this regime, where the field value is much smaller than the Planck scale,  quantum corrections to the mass from gravity are expected to be small.

In the previous work of~\cite{Abdalla:2012ug}, it was shown that the mass of the DM in this scenario would correspond to the mass of an axion-like particle for~a decay time of DE on the order of the age of the Universe, as~assumed in the work of~\cite{Abdalla:2012ug}. The~association of the equation of state of this real scalar field with the equation of state of a DM particle  can be justified by the fact that after the phase transition, the oscillations about the quadratic minimum of the potential become the main important aspect~\cite{Marsh:2015xka,Magana:2012ph}. 
Apart from providing a unified description of the dark sector, the~fact that scalar field in the MDE model has the mass of an axion-like particle, which is considered to be among the main candidates for DM today~\cite{Ferreira:2020fam, Amin:2022nlh, Nadler:2021dft, Irsic:2017yje, Dalal:2022rmp, Powell:2023jns, Semertzidis:2021rxs, Nakai:2022dni, Nakatsuka:2022gaf, QUAX:2020uxy, Cicoli:2021gss, Harigaya:2019qnl}, motivates us to further explore this model in the light of the \mbox{new constraints}~\cite{Li:2019san}.

It is important to emphasize that the field after the phase transition is not the QCD axion. Following this transition, the~oscillations around the quadratic minimum of the potential become the primary factor determining its effective equation of state. Consequently, the~field behaves in a manner consistent with a dark matter equation of state, with~a mass comparable to that of an axion-like particle.

However, despite this effective behavior \footnote{We can think of any field theory in the context of cosmology as an effective field theory valid until some energy scale. The~same is true for our model. We can think of our model as an effective model at low energies. %At higher energies this field has a different equation of state other than a dark energy oscillating fast in an effectively different potential.
 Since this field in the metastable vacuum accounts for today's dark energy, having an energy density of the order of $10^{-47}$ GeV\textsuperscript{4},  it has a negligible cosmological contribution at earlier times. Any quantum correction at early times would be associated with a field with totally  negligible contribution to the total energy, having no impact on cosmology, which is the reason why we do not explore these issues in the present work.}, the field may not possess all the characteristics of a standard axion dark matter candidate. In~fact, it is debatable whether it can be classified within the broader category of axion-like particles, which encompasses a wide range of candidates~\cite{Ferreira:2020fam}. Nonetheless, due to similarities in mass and equation of state, leading to similarities in the cosmological behavior, we will refer to this ``axion mass-like field" as an ``axion-like field" for simplicity.

Here, we revisit the MDE model using the new constraints from~\cite{Li:2019san} in~order to test if the model is still viable and whether the newly constrained decay time still predicts the same mass for the DM particle produced in this scenario.
As discussed above, due to the similarities of the current DE and the primordial inflation, one of our main goals is also to check if this model inherits the same problems of old inflation~\cite{GUTH1983321, PhysRevD.46.2384}, i.e, potentially observable inhomogeneities from the process of bubble nucleation and~evolution. 

In particular, we aim to address the following~questions:
\begin{enumerate}
\vspace{0.1cm}

\item Is MDE still a viable model considering the latest cosmological constraints?

\vspace{0.1cm}

\item What is the mass of the DM resulting from this process?

\vspace{0.1cm}

\item Would the bubble nucleation process in this model lead to inhomogeneities that could plague the model?

\vspace{0.1cm}

\item Could this model leave observational imprints that could be searched for in future experiments?

\vspace{0.1cm}
\end{enumerate}

In order to address these questions, we organize this paper as follows: In Section~\ref{sec2}, we present the MDE model proposed in~\cite{Abdalla:2012ug} to describe an interacting dark sector model based on field theory.  
In Section~\ref{sec3}, we compute the mass of the DM particle of this model that results from the DE decay with a characteristic time compatible with the recent constraints from~\cite{Li:2019san}. 
In Section~\ref{sec4}, we analyze the bubble nucleation process and its subsequent dynamics in order to see if the model would predict observable inhomogeneities. In~Section~\ref{sec5}, we conclude and discuss some future prospects. In~Appendix~\ref{app}, we demonstrate the validity of the approximations~considered.

\section{A Model for Dark Energy Decay}
\label{sec2}

In this section, we present the MDE model proposed in~\cite{Abdalla:2012ug}  to describe an interacting dark energy model based on field theory. %This provides a concrete example of the scenarios constrained in Ref.~\cite{Li:2019san} with recent data.
Since the model contemplates an energy exchange between dark energy
and dark matter, none of these components is separately conserved. The~equations describing the model are written as: %MDPI: The format of equation variables (italics, bold, superscript, uppercase, lowercase, etc.) should be consistent throughout the paper (main text, tables, figures, equations). Please double check all the cases in the paper and unify them carefully. Response: I have checked and revised all
\begin{eqnarray}
\label{rhoDE}
\dot{\rho}_{\rm DE}+3H\rho_{\rm DE}(1+\omega_{\rm DE})&=&Q_{\rm DE},\\
\label{rhoDM}
\dot{\rho}_{\rm DM}+3H\rho_{\rm DM}(1+\omega_{\rm DM})&=&Q_{\rm DM},
\end{eqnarray}
where dot denotes the derivative with respect to cosmic time, $\omega_i$ is the equation of state parameter (EoS), $H$ is the Hubble parameter, $Q_i$ is the interaction term, and the subscript $i$ indicates DE or~DM.

%Following the model proposed in Ref.~\cite{Abdalla:2012ug}, 
We consider that $Q_i$ is linearly proportional to the energy density of DE~\cite{Abdalla:2012ug}, as follows:
\begin{eqnarray}
\label{Q}
Q_{\rm DM}=-Q_{\rm DE}=\Gamma\,\rho_{\rm DE}.
\end{eqnarray}

In the model we are going to consider, the~DE decay rate, $\Gamma$, does not depend on external parameters such as the curvature of the Universe or the scale 
factor. Instead, the~DE decay rate is assumed to
be a constant (similar to the case of the radioactive decay of unstable particles and nuclei).

Once the interaction term is defined, we can  write an effective EoS for DE
\begin{eqnarray}
 \omega_{\rm DE}^{\rm (eff)}=\omega_{\rm DE}+\frac{\Gamma}{3H}.
\label{eoseff}
\end{eqnarray}

From Equation~\eqref{eoseff}, we see that in a DE-DM interaction model, the effective EoS can have a  phantom-like behavior ($\omega^{\rm (eff)}_{\rm DE}<-1$) when the decay rate is negative (energy flow  DM $\to$ DE), or~a quintessence-like behavior ($\omega^{\rm (eff)}_{\rm DE}>-1$)  when the decay rate is positive  (energy flow  DE $\to$ DM), which is our~focus.

In the MDE model, DE is represented by a scalar field with a potential energy endowed with doubly degenerate minima. 
When a small symmetry-breaking term is added, a~small energy difference between the degenerate minima emerges, which we will denote by $\epsilon$. 
The DE responsible for the current expansion of the Universe is represented by the scalar field at the metastable minima, where the potential energy is adjusted to $\epsilon\sim 10^{-47}\textrm{GeV}^4$. Such a potential can be written as
\begin{eqnarray}\label{WZ}
V(\varphi)=\left|2m\varphi-3\lambda\varphi^2\right|^2+Q(\varphi),
\end{eqnarray}
%
%\textcolor{blue}{Em vez de escrever o potencial em termos do campo carregado, podemos escrever diretamente em termos do campo real $\varphi$}
where $\varphi$ is a real scalar field of mass $m$ and coupling constant $\lambda$ and $Q(\varphi)$ is the symmetry-breaking term, which is adjusted so that we have the value of cosmological constant at the metastable minimum \footnote{Another interesting possibility is to consider the scalar field with the even self-interactions up to sixth order, as~analyzed in the work of~\cite{Landim:2016isc}.}. Of~course, this is a fine-tuned choice. This is the same fine tuning that exists in the standard $\Lambda\textrm{CDM}$ model. Although~the model analyzed here does not alleviate this fine tuning, it is an attempt to unify the dark sector by describing it through a single scalar field. In~addition, this model can be viewed as  a first step in obtaining a unified scenario  that could  describe a more dynamic dark energy. This possibility has become especially interesting  after the recent BAO data released by the Dark
Energy Spectroscopic Instrument (DESI) \cite{DESI:2024uvr, DESI:2024kob}. 

\begin{figure}[h]
    %\centering
    \includegraphics[width=0.7\textwidth]{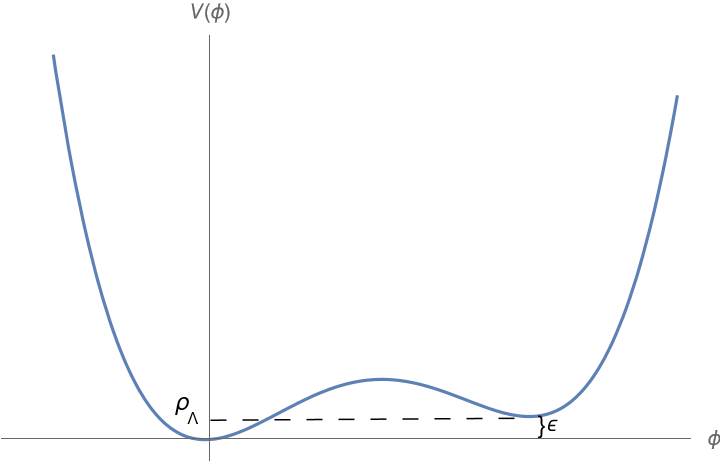}
    \caption{Shape of the potential $V(\varphi)$.}
    \label{fig:V}
\end{figure}

Equation \eqref{WZ} above is inspired in the Wess--Zumino model~\cite{osti_4338791}. The~potential is adjusted such that the stable minimum is the zero of the potential. 
The potential has minima at $\varphi=0$ and $\varphi=\frac{2m}{3\lambda}$. The~general shape of the potential is shown in Figure~\ref{fig:V}. Following the previous work~\cite{Abdalla:2012ug}, we will consider the value $\lambda=10^{-2}$ for the coupling constant. Although~this choice can seem fine tuned at first,  we will show later that the results are not very sensitive to the value of $\lambda$. In~addition,  values for the coupling term with a magnitude not much bigger or much smaller than 1 can be viewed as natural choices  in the context of Quantum Field~Theory. 
\vspace{-6pt}

In the case we are interested in, with~$\Gamma >0$ and 
  energy flowing from the cosmological constant into dark
matter, it follows that the 
value of the dark matter density parameter extrapolated to high redshifts would be higher than the value predicted by $\Lambda\textrm{CDM}$, while the expansion rate would be lower than that in $\Lambda\textrm{CDM}$. If~the energy transfer occurs at a non-negligible rate, this affects cosmological observables, which is why it is important to consider observational constraints in such~a scenario.  

In the work of~\cite{Shafieloo:2016bpk}, a generic model of metastable DE decaying into DM was analyzed and constraints on the value of the decay rate were obtained. Later in~\cite{Li:2019san}, a~further analysis of this model was performed using more recent data. In~the latter, 
 they considered four combinations of datasets in their analysis:  (1) Pantheon compilation~\cite{Pan-STARRS1:2017jku} in combination
with BAO data from 6dFGS~\cite{Beutler:2011hx}, MGS~\cite{Ross:2014qpa}, and BOSS DR12~\cite{BOSS:2016wmc}.
(2) They add BAO data from eBOSS DR14~\cite{eBOSS:2018yfg} to the first
dataset. (3) They add high redshift BAO measurement
from Ly$\alpha$ \cite{BOSS:2017uab} to the second dataset. (4) They included the CMB distance prior~\cite{Chen:2018dbv,Planck:2018vyg} to the full combination of datasets. These four analyses constrained, respectively, the following~values: (1)  $\Gamma/H_0= 0.47^{+0.5}_{-0.5}$ (mean with 1$\sigma$ deviation); (2)  $\Gamma/H_0 = 0.07^{+0.4}_{-0.3}$ ; (3)   $\Gamma/H_0=  0.51^{+0.27}_{-0.25}$; (4) $\Gamma/H_0 = -0.02^{+0.01}_{-0.01}$. More details on their analysis and methodology can be found in~\cite{Li:2019san}. 
%Recalling that the decay rate is the inverse of the decay time, this shows that the current constraints imply in a  lifetime of the metastable DE   greater than the age of the universe. 
We can  use these constraints  in order to estimate an upper limit on the decay rate and to analyze the scenario that  results from this limiting value. Although~the constraints differ depending on the dataset used, we can consider $\Gamma/H_0 \leq \mathcal{O}(10^{-1})$ as a reasonable estimate for the upper limit.  The~analysis we are going to perform in the following sections will provide us with   the mass of the 
resulting dark matter particle (Section \ref{sec3}), and~the fraction of energy currently remaining in the false vacuum (Section \ref{sec4}).

%We are going to consider the aforementioned constraints in order to compute, in the next section, the mass of the dark matter particle that will result from the phase transition in the MDE model.  %It will become clear that the results are not so sensitive to the specific limit on $\Gamma$ chosen among the results presented above. 

\section{Dark matter from first order phase transition}
\label{sec3} 

The energy transfer in the MDE model occurs  due to the tunneling from the metastable (false)
to the stable (true) vacuum of the potential, Equation~\eqref{WZ}.
This process can be described by a first-order phase transition according to the semi-classical method developed in ~\cite{Coleman:1977py, Callan:1977pt}. 
Using this framework, we are going to compute the mass of the resulting  DM particle as a function of $\Gamma$.

The decay rate (per unit volume) reads
\begin{eqnarray}
\label{Gamma}
\frac{\Gamma}{V}=\frac{S_{E}^{2}({\tilde{\varphi}}(\rho))}{(2\pi\hslash)^{2}}\times 
\left[\frac{det'(-\partial_{\mu}\partial_{\mu}+V''({\tilde{\varphi}}(\rho))}{det(-\partial_{\mu}\partial_{\mu}+V''(\varphi_{+}))}\right]^{-\frac{1}{2}} \times e^{-\left(\frac{S_{E}}{\hslash}-\frac{S_{\Lambda}}{\hslash}\right)},
\end{eqnarray}
where $\varphi_{+}$ is the field amplitude at the false vacuum and ${\tilde{\varphi}}(\rho)$ is, in~analogy with the case of
particles, the~classical field amplitude in Euclidean space crossing the potential $-V(\varphi)$ subject to boundary conditions $\varphi_{\rm initial}=\varphi_{\rm final}=\varphi_{+}$. 
$S_{E}$ is the Euclidean action 
and
$S_{\Lambda}$ is the Euclidean action of a particle at the false vacuum.
Looking for an analytical solution, we will consider the so-called thin wall approximation~\cite{Coleman:1977py, Callan:1977pt}, as conducted in the previous work of~\cite{Abdalla:2012ug}. In~this approximation, the energy difference
between the two minima of the potential, given by the parameter $\epsilon$, is considered to be small, then  perturbative results can be obtained in terms of  $\epsilon$.

The classical equation of motion in the Euclidean space for the field $\varphi$
subject to the potential $V(\varphi)$ is obtained by the minimization of $S_E$, which reads
\begin{eqnarray}
\frac{\delta S_{E}(\varphi(x))} {\delta\varphi}=-\partial_{\mu}\partial_{\mu}\varphi(x)+V'(\varphi)=0.
\end{eqnarray}

We also suppose that %MDPI: Please confirm if this paragraph no need to add indent. Response : I have checked and changed and added indent.
\begin{eqnarray}
\displaystyle\lim_{\tau\rightarrow{\pm}\infty}\varphi(\vec{x},\tau)=\varphi_{+},
\end{eqnarray}
%
%with respect to $\varphi_{\rm initial}$ and $\varphi_{\rm final}$, 
where $\tau=-it$.

The solution is Euclidean invariant, which means that 
$\varphi(\vec{x},\tau)=\varphi(({|\vec{x}|^{2}}+\tau^{2})^{\frac{1}{2}})$.
Defining $\rho\equiv(|\vec{x}|^{2}+\tau^{2})^{\frac{1}{2}}$, the~equation of motion now reads
\begin{eqnarray} \label{eqmotion}
\frac{\partial^{2}\varphi}{\partial\rho^{2}}+\frac{3}{\rho}\frac{\partial}{\partial\rho}\varphi -V'(\varphi)=0,
\end{eqnarray}
which is analogous to the equation of motion of a particle at position $\varphi$ as a function of time  $\rho$, subject to a potential
$-V(\varphi)$ and a friction~term. 

The decay process occurs by the formation of bubbles of a true vacuum surrounded by the false vacuum outside. 
The field is at rest both inside and outside, and~the friction-like term, $\frac{3}{\rho}\frac{\partial}{\partial\rho}\varphi$,  is nonzero only at the bubble wall.
Therefore, in~the case the wall is thin, we can consider that $\rho = R$ at the wall, being $R$  the bubble radius (see ~\cite{Coleman:1977py, Callan:1977pt} for further details).
Finally, for~a small energy difference $\epsilon$, the~quantity $R=\rho$ is large and the friction coefficient $1/\rho\to0$, then we can neglect the friction term also at the bubble wall.
From these considerations, Equation~\eqref{eqmotion} now reads
\vspace{-6pt}
\begin{eqnarray}
\frac{\partial^{2}\varphi}{\partial\rho^{2}}=V'(\varphi) \label{ddotphi}.
\end{eqnarray}

Then, considering the potential in  Equation~\eqref{WZ}, the~value of the field is each region is
%
%\begin{subequations}
%\begin{empheq}[left={\varphi=\empheqlbrace}]{align}
%0  & ,\quad 0 <\rho\ll R ,\\
%\tilde{\varphi}  & ,\quad \rho\approx R,\\
%2m/3\lambda  & ,\quad \rho\gg R,
%\end{empheq}
%\end{subequations}
\begin{equation}
 \varphi=
\begin{cases}
0 
   & \mbox{if } 0 < \rho\ll R\,, \\[1ex]
\tilde{\varphi} 
   & \mbox{if } R-\Delta<\rho<R+\Delta\,,
   \\[1ex]
2m/3\lambda
   & \mbox{if } \rho\gg R\,,
\end{cases}
\end{equation}
which corresponds to the regions  
inside the bubble ($0 <\rho\ll R$),
at the thin wall ($\rho\approx R$)
and
outside the bubble ($\rho\gg R$).

Now, we compute the action by adding the separated contribution of each region, which reads
\vspace{-6pt}
\begin{eqnarray}
S\equiv S_{E}-S_{\Lambda}
&\approx&  
2\pi^{2}\left(\intop_{0}^{R-\Delta}d\rho\,\rho^{3}(-\epsilon)
+
\intop_{R-\Delta}^{R+\Delta}d\rho\,\rho^{3}
\left[\frac{1}{2}\left(\frac{d\tilde{\varphi}}{d\rho}\right)^{2}+U\right] 
+\intop_{R+\Delta}^{\infty}d\rho\,\rho^{3}\, 0 \,(\textrm{GeV}^{4})\right)\,\nonumber\\
&=&-2\pi^{2}\epsilon\frac{R^{4}}{4}+2\pi^{2}R^{3}\intop_{R-\Delta}^{R+\Delta}d\rho\left(\frac{1}{2}\left(\frac{d\tilde{\varphi}}{d\rho}\right)^{2}+U\right)\nonumber\\
&=&-\frac{\pi^{2}R^{4}}{2}\epsilon+2\pi^{2}R^{3}S_{1}\label{sdeR},
\end{eqnarray}
where $\Delta$ represents the width of the wall, and~we defined
\begin{eqnarray}
\label{S1}
S_{1}\equiv \intop_{R-\Delta}^{R+\Delta}d\rho\left[\frac{1}{2}\left(\frac{d\tilde{\varphi}}{d\rho}\right)^{2}+U\right].    
\end{eqnarray}

Minimizing the action $S$ with respect to $R$, one obtains
\begin{eqnarray}
\frac{dS}{dR}=-2\pi^{2}R^{3}\epsilon+6\pi^{2}R^{2}S_{1}=0,
\label{diffaction}
\end{eqnarray}
which gives the solution
\begin{equation}\label{solutionR0}
    R=\frac{3S_{1}}{\epsilon}.
\end{equation}

For small $\epsilon$, integrating Equation~\eqref{ddotphi}, we obtain $\frac{\partial}{\partial\rho}\varphi=\sqrt{2U}$. Using this expression and considering the potential from Equation~\eqref{WZ}, one obtains from Equation~\eqref{S1}
\vspace{-6pt}
\begin{eqnarray}
\label{S1result}
S_{1}=\sqrt{2}\left(\frac{4m^{3}}{27\lambda^{2}}\right).
%\label{S1}
\end{eqnarray}

Finally, using Equation~\eqref{S1result} in Equation~\eqref{sdeR}, the~action will result in
\vspace{-6pt}
\begin{eqnarray}
S\approx\frac{m^{12}}{\lambda^{8}\epsilon^{3}}.
\end{eqnarray}

In the decay rate expression, Equation~\eqref{Gamma}, the~exponential term will dominate. We also know the pre-exponential term has the dimension of energy to the $4th$ power and that it will weight the overall value; then, we simply  estimate it as $1\, {\rm GeV}^4$ in order to provide correct units.  
Therefore, Equation~\eqref{Gamma} can be written as%~\cite{Linde:1990flp},
\begin{eqnarray}
\frac{\Gamma}{V}= \exp{\left\lbrace-\frac{m^{12}}{\lambda^{8}\epsilon^{3}}\right\rbrace} \; {\rm GeV}^{4}.
\end{eqnarray}

From the above equation, we can see that while $\lambda$ is raised to the eighth power, the~mass is raised to the twelfth power. This shows that the results we are going to obtain in this work are not very sensitive to the choice of $\lambda$.

Finally, by~substituting the values of $\epsilon$ and $\lambda$, we obtain for the decay rate (per volume)
\begin{eqnarray}\label{Gammam}
\frac{\Gamma}{V}=\exp{\left\lbrace-10^{157}\left(\frac{ m}{\rm GeV} \right)^{12}\right\rbrace} \; {\rm GeV}^{4}.
\end{eqnarray}

One can  verify that, within~the validity of the semiclassical approach  we are considering, maintaining the pre-exponential term in Equation~\eqref{Gamma}  would have changed the result at most by a factor of 2. This factor of 2 would be multiplied by a factor of $10^{157}$ in the equation above. Therefore, it is really sufficient for our purposes to consider the pre-exponential term to be of order 1 GeV\textsuperscript{4}. %We removed italic from variable folloing the equation, please confirm. Response: I confirm.

Due to the symmetry of our problem, we can invert $\Gamma/V$, take its fourth root, and interpret it as
the decay time, $t_{\rm decay}$. 
%abaixo eu achei melhor colocar a massa que foi obtida no trabalho anterior pra nao dar um destaque tao grande pra um resultado que nao eh deste trabalho, mas de um anterior. 
If we equate the decay time to the age of universe, $t_{\rm 0}=10^{17}{\rm s}$, as~was conducted in~\cite{Abdalla:2012ug}, the~ value 
$m\sim10^{-13}\,{\rm GeV}$ is obtained. 
This result implies in a mass of an axion-like DM particle~\cite{Abdalla:2012ug}. The equation of state of this real scalar field  can be associated with a DM particle since, after~the phase transition, oscillations about the quadratic minimum of the potential are the important aspect to determine the equation of state~\cite{Marsh:2015xka,Magana:2012ph}. 
%Knowing that the axion is one of the main candidates for DM today, 

We felt motivated to further explore this model in the light of the new cosmological data~\cite{Li:2019san}. 
In order to do so, let us now consider the case when the inverse decay rate is not equated to the age of the Universe, but~to an arbitrary decay time. 
We parameterize this case by introducing the parameter $\alpha$, assuming arbitrary values smaller than 1. If~we consider the quantity $(\Gamma/V)^{-1/4}$  to be related to the age of the universe $t_{0}$, the~quantity $(\alpha \; \Gamma/V)^{-1/4}$ will imply in a decay time   of the form $t_{\rm 0}/(\alpha)^{1/4}$. %$t_{\rm decay}/(\alpha)^{1/4}$.
As a consequence, different choices of the parameter $\alpha$ (different choices of the decay rate/decay time) result in different values for the mass $m$ of the  DM~particle. 

\begin{figure}[h]
%\centering 
\includegraphics[width=0.7\textwidth]{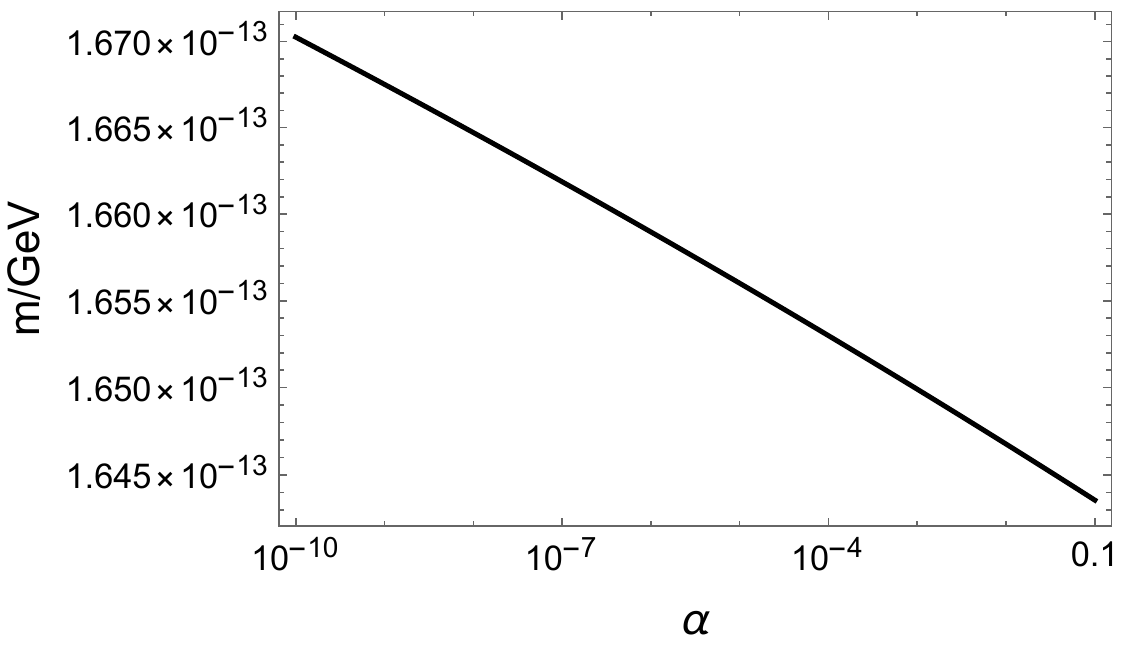} 
\caption{Dependence of the mass on the parameter $\alpha$, with~the latter ranging from $10^{-10}$ to $10^{-1}$.}
\label{malpha}
\end{figure}

In Figure~\ref{malpha}, we illustrate the dependence of the mass of the resulting DM particle on $\alpha$, with~$\alpha$ ranging from $10^{-10}$ to $10^{-1}$.  
From this figure, we notice that there is no change in the order of magnitude of the mass for this whole range of values of $\alpha$.

The behavior shown in Figure~\ref{malpha} can be described by the analytical expression

\begin{eqnarray}
m(\alpha)=1.12246\times10^{-13}
   \sqrt[12]{94.8245-\ln\,\alpha},    
\end{eqnarray}
which shows the weak dependence of the mass on $\alpha$. The~order of magnitude of the mass only changes for $\alpha\sim 10^{-107598}$, which corresponds to a decay time $10^{26900}t_{\rm 0}$. 
Obviously, such a case is not interesting for us. 
Therefore, for~any non-negligible (and observationally allowed) decay rate, the~prediction for the DM mass, $m\sim10^{-13}\,{\rm GeV}$, remains consistent. In~addition to that, one can verify that these results also do not change if one considers other similar values for the coupling constant, as~$\lambda=10^{-1}$ for~instance.

As is well known, first-order phase transitions occur through the random nucleation of bubbles. 
In this process, firstly the false vacuum energy is transferred to the kinetic energy of the bubbles wall, which asymptotically expands at the speed of light~\cite{Coleman:1977py, TeppaPannia:2016hwv, Simon:2009nb, Fischler:2007sz}.   
However, in~some cases, at~some point, the~walls of these bubbles may percolate, and~the energy of the walls is converted into particles (energy density) that eventually thermalize~\cite{hawking1982bubble, PhysRevD.46.2384,kosowsky1992gravitational, PhysRevD.84.024006}. Let us analyze this process further in the following~section.

\section{Bubble Nucleation}
\label{sec4}

First-order phase transitions, as~the one considered here, occur through the nucleation and growth of bubbles of the new phase~\footnote{Evolution of bubbles of new vacuum in de Sitter backgrounds has been studied in different context (see, for~instance,~\mbox{\cite{Pannia:2021lso, TeppaPannia:2016hwv, Aguirre:2009ug, Simon:2009nb, Fischler:2007sz, PhysRevD.84.024006}}). In~addition, first-order phase transitions have recently been considered as one of the possible explanations for the positive evidence  of a low-frequency stochastic gravitational-wave  background found in PTA experiments, see for example~\cite{NANOGrav:2023hvm}. Another recent interesting  application of  first-order transitions is the New Early dark Energy models, see for instance~\cite{Niedermann:2019olb}.}.
This is similar to the process that happens in old inflationary models. In~the case of old inflation, in~order for the decay process to occur efficiently, it was necessary to have an inflation decay rate  higher than a certain minimum value, in~order to end inflation successfully. However, such a high decay rate would imply a bubble nucleation process, which leads to a highly inhomogeneous Universe. Given the similarities between the phase transition described here and the process that plagued the old inflationary model, we think it is important to discuss some implications of such a process in the context of the model  considered~here. 

In our low-temperature late-time  model for the dark sector, the~bubble nucleation occurs through Coleman--Callan  tunneling and the nucleation rate is essentially temperature-independent. Therefore, we consider the idealized model for the transition described in~\cite{GUTH1983321}, in~which the Universe is taken to be always at zero temperature, with~negligible curvature. We can also approximately consider a de Sitter exponential expansion, $a(t) \propto e^{H\, t}$, in~the late time Universe, $H$ being the Hubble parameter assumed to be approximately constant. This is a sufficiently good approximation for our practical purposes. The~de Sitter approximation  can also be justified by the fact that the recent constraints that we are considering for the  decay time implies a decay process that will only be effective in a distant future when the Universe is even more close to a de Sitter~expansion. 

We consider that bubble nucleation begins at a time $t_{B}$, and~afterward occurs at a constant rate per unit of physical volume $\Gamma/V_{0}$. 
The decay rate per unit of coordinate volume is then given by $\Gamma(t)/V=(\Gamma/V_{0})\textrm{e}^{3H t}$. We can consider  the approximation in which a bubble starts expanding from a very small radius $r_{0}\ll\,H^{-1}$. In~order to show that this is the case, 
let us estimate the initial radius of the bubble from which it starts expanding. Using Equations~\eqref{solutionR0} and~\eqref{S1result}, we can estimate this radius to be
\begin{equation}
\label{R0}
R_{0}=\sqrt{2}\left(\frac{12m^{3}}{27\lambda^{2}\epsilon}\right) \approx 10^{-2}\textrm{cm},
\end{equation}
where in the last equality we used the particle mass determined in Section~\ref{sec3}, $m \approx 10^{-13}{\rm GeV}$, together with the values  $\lambda=10^{-2}$ and $\epsilon=10^{-47}{\rm GeV}^{4}$.

%TURNER pg 46 segundo e terceiro paragrafo. Why we are interested in big bubbles:

In a successful first-order phase transition, most of the bubbles are nucleated and collide
in a time interval comparable to or shorter than the Hubble time. This is the so-called fast transition, in~which the nucleation rate is comparable to the expansion rate of the Universe~\cite{GUTH1983321, PhysRevD.46.2384}. After~bubbles collide, in~successful cases, the~distribution of energy in the Universe becomes homogeneous as the relevant bubbles are sub-horizon sized
when they collide. In~the slow transitions, the nucleation rate is much smaller than the expansion rate of the Universe. We will show that in the case of the MDE model considered here, rare and very large bubbles are nucleated in a cosmological time.%Please ensure original meaning retained. 
 These bubbles can grow to
astrophysical sizes and their dynamics cover a much longer period of time. After~such a bubble is nucleated, it soon starts expanding at the speed of light~\cite{Coleman:1977py, Callan:1977pt}. For~a bubble that emerges at a point $(t_B,\Vec{x}_0)$ with a very small radius, at~a later time $t$ its radius will be given by
\vspace{-6pt}
\begin{equation}\label{r}
    r(t,t_{B})=\left|\Vec{x}-\Vec{x}_{0}\right|=\int_{t_B}^{t} dt'a(t')^{-1} \;.
\end{equation}

Evaluating the integral above gives us the result $r(t,t_{B})=H^{-1}(\textrm{e}^{-H t_{B}}-\textrm{e}^{-H t})$. 
We can see that the bubble radius will asymptotically assume the finite value
\begin{equation}\label{limitr}
r_{A}\equiv \lim_{t \rightarrow \infty}    r(t,t_{B})=H^{-1}\textrm{e}^{-H t_B},
\end{equation}
hence, the~volume of the Universe occupied by this bubble will asymptotically  be
\begin{equation}
    V_\textrm{asym}=\frac{4 \pi}{3 H^3}\textrm{e}^{-3H t_B} \;.
\end{equation}

Due to the approximately de Sitter expansion, the~bubble will grow in comoving size for only about a $H^{-1}$ time, and~after that, being super-Hubble, it simply stretches with the scale factor as the universe expands. This suggests that independently from the time past after the nucleation, two bubbles that emerge simultaneously at points $(t_B,\Vec{x}_0')$ and $(t_B,\Vec{x}_0'')$ at a proper distance $D=a(t_B)\times\left|\Vec{x}_0'-\Vec{x}_0''\right|>2H^{-1}$ will never~percolate.

%It is possible to analyze the evolution of the  bubble distribution in the universe by the parameter $p(t)$, the probability that an arbitrary point remains in the false vacuum phase. Since we are considering the random nucleation of bubbles, i.e.,~the bubbles are totally uncorrelated, except by the exclusion principle that bubbles do not form inside of bubbles, the probability $p(t)$ is given by~\cite{Guth:1979bh, Guth:1981uk, Guth:1980zm, PhysRevD.46.2384, GUTH1983321}:
It is possible to analyze the evolution of the  bubble distribution in the Universe by investigating the physical volume remaining in the false vacuum, given by $V_{phys} \propto a^{3}(t)p(t)$, where $p(t)$ is the probability that a given point in space is in the false vacuum at a  time $t$. The quantity  $p(t)$ can be written in the following form~\cite{Guth:1979bh, Guth:1981uk, Guth:1980zm, PhysRevD.46.2384, GUTH1983321}
\begin{equation}
   p(t)=\textrm{e}^{-I(t)} \;,
\label{p(t)}
\end{equation}
where $I(t)$ is the expected volume of true-vacuum bubbles per unit volume of space at time $t$ \footnote{As discussed in~\cite{PhysRevD.46.2384},  the~exponentiation of $I(t)$ corrects for some effects, like the fact that when calculating $I(t)$, regions in which bubbles overlap are counted twice. Furthermore, the virtual bubbles which would have nucleated and had their point of nucleation not already be in a true-vacuum region are also included.}. For a  phase transition beginning at time $t_0$, we can expect this volume to be given, at~time $t$, by~the expression
\begin{equation}
    I(t)=\frac{4 \pi}{3}\int^{t}_{t_0} dt' (\Gamma/V_{0})a^3(t')r^3(t,t') \;,
\label{i(t)}
\end{equation}
where a constant decay rate was  considered  $\Gamma$ for the  MDE model. Above,  $r(t,t')$  is the coordinate
radius at time t of a bubble that was nucleated at a time $t'$, expressed in Equation~\eqref{r}. Note that we are integrating in the nucleation times. Above,~we considered that the decay rate in the MDE model is a~constant.

%Since the half-life of the DE will be many times the age of the universe, we can consider that $H(t-t_{B})\gg 1$. Then $I(t)$ will be 

%\begin{equation}
 %   I(t)=\frac{4 \pi}{3} \frac{\Gamma}{H^{3}} \textrm{e}^{3H t} \;.
  %  \label{i(t)solution}
%\end{equation}

%Combining Equation~(\ref{i(t)solution}) with Equation~(\ref{p(t)}), 

%Abaixo seguirei a abordagem do Turner Equation~(4).4 e discussao que se segue, para poder conectar as eqs de p e I com a quantidade que importa e que iremos analisar quantitativamente que eh (Gamma/V0)/H^4
The equation above can be better understood if we view the quantity $I(t)$   as a function of the
scale factor  rather than  $t$. In~this case, we can rewrite the equation above as
\begin{equation}
    I(t)=\frac{4 \pi}{3}\int^{R(t)}_{R_0} da \frac{(\Gamma/V_{0})}{H}a^2\left(\int^{R(t)}_{a} da' \frac{1}{a'^{2}H} \right)^{3} \;.
\label{i(t)2}
\end{equation}

Above, we used Equation~\eqref{r} for $r(t,t')$.  The~integrals in this expression are dominated by the upper limit of the integration range. We can see that the quantity $(\Gamma/V_{0})/H^{4}$ sets the
magnitude of $I(t)$. Up~to a numerical factor, this quantity measures the fraction of space occupied by large bubbles, which  is given by $1-p(t)$. Therefore, provided that this fraction is small, as~it will be in all cases of interest,  $1-p(t) \approx I(t)$.

% fim da pg 40 (ou 2 no pdf) de:
%Some problems with extended inflation-Weinberg

%https://sci-hub.wf/10.1103/physrevd.40.3950

Even expanding at the speed of light, a~bubble which nucleates at a time $t_{B}$ can only grow to a finite comoving radius given by Equation~\eqref{limitr} . Then, as~shown above, if~the
separation between two bubbles at time $t$ is greater than 2$r_{A}$, the~bubbles will never meet. Therefore, the bubbles nucleated in a time interval of duration of~$H^{-1}$ can never fill space by themselves, but~instead only occupy a fraction of order $(\Gamma/V_{0})/H^{4}$ of the region which remained in the old phase at the time that they were nucleated.  Although~bubble nucleation continues indefinitely, and~despite that the physical volume of the old phase region (proportional to $a(t)^{3}p(t)$) is
an increasing function of time, the~bubbles produced have smaller and smaller comoving volume and so can fit in the remaining regions of old phase without~overlapping. 

Another simple way of understanding this is remembering that after a bubble is nucleated and  grown to a size of about a Hubble radius, its size simply conformally stretches as the Universe expands. Then, from that point on, the~volume fraction of the Universe that it occupies remains constant. The~volume fraction occupied by the bubbles nucleated during the time interval $\Delta t = H^{-1}$ is roughly the volume of  a bubble when it begins conformally stretching, which is around $H^{-3}$, times the number of such bubbles nucleated in this interval per unit volume given by $\Gamma/(V_{0}H)$. Therefore the quantity $(\Gamma/V_{0})/ H^{4}$ indicates  the volume fraction of space occupied by bubbles nucleated over a Hubble time at a given epoch. We will denote by $\xi$ this important quantity,
\begin{equation}
    \xi \equiv \frac{\Gamma/V_{0}}{H^{4}}.
\end{equation}

A slow transition is considered to be the one in which the quantity above is much smaller than one, $\xi \ll 1$, and~the Coleman--Callan tunneling is the only significant mechanism of bubble nucleation. When this is not the case, the~transition is considered to be~fast.

The decay time in our model is given by the quantity $t_{\rm decay}=(\Gamma/V)^{-1/4}$. As~discussed in Section~\ref{sec3}, we can write this decay time as $t_{decay}\equiv \alpha^{-1/4} t_{0}$. Therefore, we can write the quantity $\xi$  as  \begin{equation}\label{xi_bubble}
    \xi =\frac{t_{\rm decay}^{-4}}{t_{0}^{-4}} = \alpha.
\end{equation}
%where in the second equality we considered the limit $t_{\rm decay} \gtrsim 10\, t_{0}$. %We can see from the above equation that $\xi\ll 1$, proving that we are in the regime of slow transition.

%ACIMA, REFERIR AOS VALORES DE gAMMA/h0 DA TABELA 4 DE https://arxiv.org/pdf/1904.03790.-> feito, conferir 

We can consider in the above equation the absolute mean values for $\Gamma/H_{0}$ described in Section II, which were obtained  in the analysis of~\cite{Li:2019san} (see Table~4 of this reference). For~the values constrained with any of the four datasets used, Equation \eqref{xi_bubble} gives us $\xi\ll 1$. Therefore, one can  safely consider the slow transition~regime.

 %We can consider in the above equation the best fit values obtained for $\Gamma/H_{0}$  in Ref.~\cite{Li:2019san} (see Table~4 of this reference). Considering the constraints that were obtained  using any of the more complete datasets 2, 3 and 4  (which in addition to Pantheon + 6dFGS + MGS + BOSS DR12,  include also BOSS DR14,  Ly$\alpha$ and  CMB, respectively),  we obtain $\xi\ll 1$. Therefore, one can  safely consider the slow transition regime\footnote{Considering in  Equation~(\ref{xi_bubble}) the constraints on $\Gamma/H_{0}$ obtained in~\cite{Li:2019san} for the dataset 1 (only Pantheon + 6dFGS + MGS + BOSS DR12), one obtain $\xi < 1$.}.

%ver slide 8 de 

%https://www.mpi-hd.mpg.de/lin/seminar_theory/talks/Talk_Winkler_280621.pdf

We can hence conclude that  such a model will not drive a complete transition to a dark matter-dominated phase. Even in the future,  there will be a dominant region of the universe in the metastable phase, where we  would still have an approximate de Sitter expansion. However, unlike old inflation, there is no observational restriction imposing that a complete transition from the dark energy state into dark matter must take~place.

As shown in this section, the~initial radius of the bubble from which it starts  expanding is $R_{0}\approx 10^{-2}$ cm. Although~it  expands very fast, the~volume fraction of space occupied by bubbles nucleated over a Hubble time at a given epoch is much smaller than one, as~shown above. Therefore, there are negligible cosmological consequences from the energy density associated with these bubbles. Since here we are only interested in the cosmological scales,  we will not  discuss further the issue of domain walls or gravitational waves.

\section{Conclusion  and Prospects}
\label{sec5}

We considered the model proposed in~\cite{Abdalla:2012ug}, in~which an energy transfer from dark energy into dark matter is described
in field theory by a first-order phase transition. We further investigate this model in light of the recent cosmological data. We find that the model is not excluded by the data, although~it is not distinguishable from $\Lambda\textrm{CDM}$. Since recent data constrain a decay time for metastable dark energy considerably larger than the age of the Universe, there is currently no prospect of observing the outcome of the DE decay process, which we show to be a dark matter particle with a mass corresponding to that of an axion-like particle,  $m \sim 10^{-13}{\rm GeV}$. %Please ensure original meaning retained. Response: I confirm.
 We analyzed, for~the first time, the~process of bubble nucleation in this model, showing  that such a model would not drive a complete transition to a dark matter-dominated phase even in a distant~future.

In particular, in~this work, we  provided the following answers to some, until~now, open questions:

\begin{enumerate}

\vspace{0.1cm}

\item Considering the recent cosmological data, the~model proposed in~\cite{Abdalla:2012ug} can still be considered, formally, a~viable model for describing an unified dark~sector.

\vspace{0.1cm}

\item The recent constraints in the decay time of the metastable dark energy  imply in a   resulting DM with a mass of an axion-like particle, although~this resulting DM would only appear in the far~future.

\vspace{0.1cm}

\item We do not expect this model to lead to observational imprints that could be searched for in future experiments, unless~extra couplings are added to the Lagrangian of the~model.

\vspace{0.1cm}

\item The bubble nucleation process was analyzed and we showed that the model considered, apart from~not leading to current observable inhomogeneities, would not drive a complete transition to a dark matter-dominated phase, even in the far~future.

\vspace{0.1cm}
\end{enumerate}

 We can conclude that in order for this   model to successfully describe a  transition to the true vacuum, the~required value of the DE decay rate would need to be bigger than the range allowed by the current observations. Although~this model does not currently inherit the characteristics of a typical interacting dark sector model due to the large decay time, it still presents some qualitative advantages, as~it is able to describe a dark sector in a unified manner through a single scalar field. %In addition,  the~fact that around the true vacuum, this field behaves as a  DM  consistent with one of the best motivated DM candidates, gives us an indication of the potential of the model. 
In addition,  the~fact that around the true vacuum, this field behaves as a  DM  with a mass consistent with the one of an axion-like particle, gives us an indication of the potential of the model.
For these reasons, we believe that possible extensions of this model deserve further investigation, as~they could lead to potentially observable signatures in case additional couplings are included in the model Lagrangian (in this context see for example the works of~\cite{Landim:2016isc, Landim:2017lyq}). The~model analyzed here  can be viewed as  a first step in obtaining a unified scenario  that could  describe a more dynamic dark energy.  
Furthermore, recently the new BAO data released by the Dark
Energy Spectroscopic Instrument (DESI)  \cite{DESI:2024uvr, DESI:2024kob} have encouraged the analysis of more dynamical DE models. In~addition to that, there is evidence from different theoretical contexts that exact de Sitter solutions with a positive cosmological constant
may not be suitable to describe the late-time universe. DE models based on scalar fields evolving in time are more promising in this regard, although~in the context of the Swampland conjecture, for example, they still have to satisfy certain criteria~\cite{Palti:2019pca, Heisenberg:2018rdu, Heisenberg:2018yae}. Verifying whether extensions of the model here could satisfy these and other theoretical conjectures, and~leave observable traces, is an issue left for a future~work. 

% como perspectiva tvz mencionar tb considerar a prossibilidade de uma barreira variável no tempo (ver arxiv 1104.4791 cosmological consequences of nearly conformal dynamics on tev scale)

\appendix

\section{Gravitational Effects}
\label{app}
In this Appendix we will discuss whether  it is  necessary to include corrections in the action due to gravitational effects. In order to investigate these effects, let us  work with the following action,
\begin{equation}
\bar{S}=\int d^{4}x\sqrt{-g}\left(\frac{1}{2}g^{\mu\nu}\partial_{\mu}\varphi\partial_{\nu}\varphi-V(\varphi)-\frac{ \mathcal{R}}{16\pi G}\right) \;,
\label{modifiedaction}
\end{equation}
where $ \mathcal{R}$ is the curvature scalar. Using the thin wall approximation, which is still valid for the action~\eqref{modifiedaction} in the cases of interest, its straightforward to show the following  relation between the action $\bar{S}$ and the action $S_0$ without gravitational effects~\cite{Coleman:1980aw},

\begin{equation}
    \Bar{S}=\frac{S_0}{\left[1+\left(\frac{R_0}{2\Delta}\right)^2\right]^2} \;.
\label{modifiedactionrelation}
\end{equation}

Above $R_0$ is the initial radius of the bubble calculated in Eq.~\eqref{R0} and $\Delta$ is
the Schwarzschild radius associated to the bubble of new
vacuum, which is given by $\Delta=2 G \epsilon (4\pi R_{0}^{3}/3)$. The expression for $\Delta$ can be understood from the fact that  in the decay process from the metastable vacuum with energy $\epsilon$ to the stable vacuum with zero energy, there exists liberation of energy proportional to the energy density of the metastable vacuum and the volume of the nucleated new vacuum bubble. From Eq.~\eqref{modifiedactionrelation} we can see that if $R_0/\Delta\sim 1$ we need to consider gravitational effects in our calculations. The gravitational effects will be important when the radius of the new vacuum bubble is of the order of the Schwarzschild radius.
 We can obtain the radius of a nucleated bubble  that would be is equal to its Schwarzschild radius by equating $R_{0}=\Delta$, which gives us

\begin{equation}
    R_{0}=\sqrt{\frac{3}{8\pi G \epsilon}}\;.
    \label{bubbleschwarzchildradius}
\end{equation}

Using Eqs.~\eqref{sdeR} and~\eqref{S1} we can see that such initial radius would  correspond to the decay of a particle with a mass of order $m \sim 10^{-2,8}\,\textrm{GeV}$, which is much heavier than the particle we obtained. Therefore the addition of gravitational effects would not significantly alter the results we obtained.

\begin{acknowledgments}
J.S.T.S is supported by the Funda\c{c}\~{a}o Coordena\c{c}\~{a}o de Aperfei\c{c}oamento de Pessoal de Nível Superior (CAPES).
L.L.G is supported by research grants from Conselho Nacional de
Desenvolvimento Cient\'{\i}fico e Tecnol\'ogico (CNPq), Grant
No. 307636/2023-2 and from the
Fundacao Carlos Chagas Filho de Amparo a Pesquisa do Estado do Rio de
Janeiro (FAPERJ), Grant No.  E-26/201.297/2021. 
L.L.G also  thank   
 Prof. Elisa Ferreira for the important discussions.
\end{acknowledgments}

\bibliographystyle{unsrt}
\bibliography{Bibilography.bib}

\begin{thebibliography}{10}

\bibitem{Sahni:2004ai}
Varun Sahni.
\newblock {Dark matter and dark energy}.
\newblock {\em Lect. Notes Phys.}, 653:141--180, 2004.

\bibitem{Ferreira:2020fam}
Elisa G.~M. Ferreira.
\newblock {Ultra-light dark matter}.
\newblock {\em Astron. Astrophys. Rev.}, 29(1):7, 2021.

\bibitem{Amin:2022nlh}
Mustafa~A. Amin and Mehrdad Mirbabayi.
\newblock {A lower bound on dark matter mass}.
\newblock 11 2022.

\bibitem{Nadler:2021dft}
Ethan~O. Nadler, Simon Birrer, Daniel Gilman, Risa~H. Wechsler, Xiaolong Du,
  Andrew Benson, Anna~M. Nierenberg, and Tommaso Treu.
\newblock {Dark Matter Constraints from a Unified Analysis of Strong
  Gravitational Lenses and Milky Way Satellite Galaxies}.
\newblock {\em Astrophys. J.}, 917(1):7, 2021.

\bibitem{Irsic:2017yje}
Vid Ir\v{s}i\v{c}, Matteo Viel, Martin~G. Haehnelt, James~S. Bolton, and
  George~D. Becker.
\newblock {First constraints on fuzzy dark matter from Lyman-$\alpha$ forest
  data and hydrodynamical simulations}.
\newblock {\em Phys. Rev. Lett.}, 119(3):031302, 2017.

\bibitem{Dalal:2022rmp}
Neal Dalal and Andrey Kravtsov.
\newblock {Excluding fuzzy dark matter with sizes and stellar kinematics of
  ultrafaint dwarf galaxies}.
\newblock {\em Phys. Rev. D}, 106(6):063517, 2022.

\bibitem{Powell:2023jns}
Devon~M. Powell, Simona Vegetti, J.~P. McKean, Simon D.~M. White, Elisa G.~M.
  Ferreira, Simon May, and Cristiana Spingola.
\newblock {A lensed radio jet at milli-arcsecond resolution II: Constraints on
  fuzzy dark matter from an extended gravitational arc}.
\newblock 2 2023.

\bibitem{Semertzidis:2021rxs}
Yannis~K. Semertzidis and SungWoo Youn.
\newblock {Axion dark matter: How to see it?}
\newblock {\em Sci. Adv.}, 8(8):abm9928, 2022.

\bibitem{Nakai:2022dni}
Yuichiro Nakai, Ryo Namba, and Ippei Obata.
\newblock {Peaky Production of Light Dark Photon Dark Matter}.
\newblock 12 2022.

\bibitem{Nakatsuka:2022gaf}
Hiromasa Nakatsuka, Soichiro Morisaki, Tomohiro Fujita, Jun'ya Kume, Yuta
  Michimura, Koji Nagano, and Ippei Obata.
\newblock {Stochastic effects on observation of ultralight bosonic dark
  matter}.
\newblock 5 2022.

\bibitem{QUAX:2020uxy}
D.~Alesini et~al.
\newblock {Realization of a high quality factor resonator with hollow
  dielectric cylinders for axion searches}.
\newblock {\em Nucl. Instrum. Meth. A}, 985:164641, 2021.

\bibitem{d28dd6a75fda4c2eb0aa9e85b7da702e}
{Sean M.} Carroll, {William H.} Press, and {Edwin L.} Turner.
\newblock The cosmological constant.
\newblock {\em Annual Review of Astronomy and Astrophysics}, 30(1):499--542,
  1992.

\bibitem{Martin:2012bt}
Jerome Martin.
\newblock {Everything You Always Wanted To Know About The Cosmological Constant
  Problem (But Were Afraid To Ask)}.
\newblock {\em Comptes Rendus Physique}, 13:566--665, 2012.

\bibitem{Palti:2019pca}
Eran Palti.
\newblock {The Swampland: Introduction and Review}.
\newblock {\em Fortsch. Phys.}, 67(6):1900037, 2019.

\bibitem{Heisenberg:2018rdu}
Lavinia Heisenberg, Matthias Bartelmann, Robert Brandenberger, and Alexandre
  Refregier.
\newblock {Dark Energy in the Swampland II}.
\newblock {\em Sci. China Phys. Mech. Astron.}, 62(9):990421, 2019.

\bibitem{Heisenberg:2018yae}
Lavinia Heisenberg, Matthias Bartelmann, Robert Brandenberger, and Alexandre
  Refregier.
\newblock {Dark Energy in the Swampland}.
\newblock {\em Phys. Rev. D}, 98(12):123502, 2018.

\bibitem{Polyakov:2007mm}
A.~M. Polyakov.
\newblock {De Sitter space and eternity}.
\newblock {\em Nucl. Phys. B}, 797:199--217, 2008.

\bibitem{Polyakov:2012uc}
A.~M. Polyakov.
\newblock {Infrared instability of the de Sitter space}.
\newblock 9 2012.

\bibitem{Valiviita:2008iv}
Jussi Valiviita, Elisabetta Majerotto, and Roy Maartens.
\newblock {Instability in interacting dark energy and dark matter fluids}.
\newblock {\em JCAP}, 07:020, 2008.

\bibitem{Mazur:1986et}
Pawel Mazur and Emil Mottola.
\newblock {Spontaneous Breaking of De Sitter Symmetry by Radiative Effects}.
\newblock {\em Nucl. Phys. B}, 278:694--720, 1986.

\bibitem{Mottola:1985ee}
Emil Mottola.
\newblock {A Quantum Fluctuation Dissipation Theorem for General Relativity}.
\newblock {\em Phys. Rev. D}, 33:2136, 1986.

\bibitem{Brandenberger:2018fdd}
Robert Brandenberger, Leila~L. Graef, Giovanni Marozzi, and Gian~Paolo Vacca.
\newblock {Backreaction of super-Hubble cosmological perturbations beyond
  perturbation theory}.
\newblock {\em Phys. Rev. D}, 98(10):103523, 2018.

\bibitem{Abramo:1997hu}
L.~Raul~W. Abramo, Robert~H. Brandenberger, and Viatcheslav~F. Mukhanov.
\newblock {The Energy - momentum tensor for cosmological perturbations}.
\newblock {\em Phys. Rev. D}, 56:3248--3257, 1997.

\bibitem{Mukhanov:1996ak}
Viatcheslav~F. Mukhanov, L.~Raul~W. Abramo, and Robert~H. Brandenberger.
\newblock {On the Back reaction problem for gravitational perturbations}.
\newblock {\em Phys. Rev. Lett.}, 78:1624--1627, 1997.

\bibitem{Finelli:2001bn}
F.~Finelli, G.~Marozzi, G.~P. Vacca, and Giovanni Venturi.
\newblock {Energy momentum tensor of field fluctuations in massive chaotic
  inflation}.
\newblock {\em Phys. Rev. D}, 65:103521, 2002.

\bibitem{Finelli:2003bp}
F.~Finelli, G.~Marozzi, G.~P. Vacca, and Giovanni Venturi.
\newblock {Energy momentum tensor of cosmological fluctuations during
  inflation}.
\newblock {\em Phys. Rev. D}, 69:123508, 2004.

\bibitem{Marozzi:2006ky}
G.~Marozzi.
\newblock {Back-reaction of Cosmological Fluctuations during Power-Law
  Inflation}.
\newblock {\em Phys. Rev. D}, 76:043504, 2007.

\bibitem{Brandenberger:1999su}
Robert~H. Brandenberger.
\newblock {Back reaction of cosmological perturbations}.
\newblock In {\em {3rd International Conference on Particle Physics and the
  Early Universe}}, pages 198--206, 2000.

\bibitem{deSa:2022hsh}
Ramon de~S\'a, Micol Benetti, and Leila~Lobato Graef.
\newblock {An empirical investigation into cosmological tensions}.
\newblock {\em Eur. Phys. J. Plus}, 137(10):1129, 2022.

\bibitem{DiValentino:2019exe}
Eleonora Di~Valentino, Ricardo~Z. Ferreira, Luca Visinelli, and Ulf Danielsson.
\newblock {Late time transitions in the quintessence field and the $H_0$
  tension}.
\newblock {\em Phys. Dark Univ.}, 26:100385, 2019.

\bibitem{DiValentino:2019ffd}
Eleonora Di~Valentino, Alessandro Melchiorri, Olga Mena, and Sunny Vagnozzi.
\newblock {Interacting dark energy in the early 2020s: A promising solution to
  the $H_0$ and cosmic shear tensions}.
\newblock {\em Phys. Dark Univ.}, 30:100666, 2020.

\bibitem{Zhao:2017cud}
Gong-Bo Zhao et~al.
\newblock {Dynamical dark energy in light of the latest observations}.
\newblock {\em Nature Astron.}, 1(9):627--632, 2017.

\bibitem{Weinberg:2000yb}
Steven Weinberg.
\newblock {The Cosmological constant problems}.
\newblock In {\em {4th International Symposium on Sources and Detection of Dark
  Matter in the Universe (DM 2000)}}, pages 18--26, 2 2000.

\bibitem{Ferreira:2014jhn}
Elisa G.~M. Ferreira, Jerome Quintin, Andre~A. Costa, E.~Abdalla, and Bin Wang.
\newblock {Evidence for interacting dark energy from BOSS}.
\newblock {\em Phys. Rev. D}, 95(4):043520, 2017.

\bibitem{Amendola:1999er}
Luca Amendola.
\newblock {Coupled quintessence}.
\newblock {\em Phys. Rev. D}, 62:043511, 2000.

\bibitem{Abdalla:2009mt}
Elcio Abdalla, L.~Raul Abramo, and Jose C.~C. de~Souza.
\newblock {Signature of the interaction between dark energy and dark matter in
  observations}.
\newblock {\em Phys. Rev. D}, 82:023508, 2010.

\bibitem{Faraoni:2014vra}
Valerio Faraoni, James~B. Dent, and Emmanuel~N. Saridakis.
\newblock {Covariantizing the interaction between dark energy and dark matter}.
\newblock {\em Phys. Rev. D}, 90(6):063510, 2014.

\bibitem{He:2008tn}
Jian-Hua He and Bin Wang.
\newblock {Effects of the interaction between dark energy and dark matter on
  cosmological parameters}.
\newblock {\em JCAP}, 06:010, 2008.

\bibitem{Costa:2013sva}
Andr\'e~A. Costa, Xiao-Dong Xu, Bin Wang, Elisa G.~M. Ferreira, and E.~Abdalla.
\newblock {Testing the Interaction between Dark Energy and Dark Matter with
  Planck Data}.
\newblock {\em Phys. Rev. D}, 89(10):103531, 2014.

\bibitem{Benetti:2021div}
Micol Benetti, Humberto Borges, Cassio Pigozzo, Saulo Carneiro, and Jailson
  Alcaniz.
\newblock {Dark sector interactions and the curvature of the universe in light
  of Planck's 2018 data}.
\newblock {\em JCAP}, 08:014, 2021.

\bibitem{Li:2019san}
Xiaolei Li, Arman Shafieloo, Varun Sahni, and Alexei~A. Starobinsky.
\newblock {Revisiting Metastable Dark Energy and Tensions in the Estimation of
  Cosmological Parameters}.
\newblock {\em Astrophys. J.}, 887:153, 4 2019.

\bibitem{Urbanowski:2021waa}
Krzysztof Urbanowski.
\newblock {Cosmological \textquotedblleft{}constant\textquotedblright{} in a
  universe born in the metastable false vacuum state}.
\newblock {\em Eur. Phys. J. C}, 82(3):242, 2022.

\bibitem{Urbanowski:2022iug}
K.~Urbanowski.
\newblock {A universe born in a metastable false vacuum state needs not die}.
\newblock {\em Eur. Phys. J. C}, 83(1):55, 2023.

\bibitem{Landim:2016isc}
Ricardo~G. Landim and Elcio Abdalla.
\newblock {Metastable dark energy}.
\newblock {\em Phys. Lett. B}, 764:271--276, 2017.

\bibitem{Landim:2017lyq}
Ricardo~G. Landim, Rafael J.~F. Marcondes, Fabr\'\i{}zio~F. Bernardi, and Elcio
  Abdalla.
\newblock {Interacting Dark Energy in the Dark $SU(2)_{R}$ Model}.
\newblock {\em Braz. J. Phys.}, 48(4):364--369, 2018.

\bibitem{Stojkovic:2007dw}
Dejan Stojkovic, Glenn~D. Starkman, and Reijiro Matsuo.
\newblock {Dark energy, the colored anti-de Sitter vacuum, and LHC
  phenomenology}.
\newblock {\em Phys. Rev. D}, 77:063006, 2008.

\bibitem{Greenwood:2008qp}
Eric Greenwood, Evan Halstead, Robert Poltis, and Dejan Stojkovic.
\newblock {Dark energy, the electroweak vacua and collider phenomenology}.
\newblock {\em Phys. Rev. D}, 79:103003, 2009.

\bibitem{Abdalla:2012ug}
Elcio Abdalla, L.~L. Graef, and Bin Wang.
\newblock {A Model for Dark Energy decay}.
\newblock {\em Phys. Lett. B}, 726:786--790, 2013.

\bibitem{Casey:2024jep}
Richard Casey and Cosmin Ilie.
\newblock {Dark Sector Tunneling Field Potentials for a Dark Big Bang}.
\newblock 2407.05752 2024.

\bibitem{Freese:2023fcr}
Katherine Freese and Martin~Wolfgang Winkler.
\newblock {Dark matter and gravitational waves from a dark big bang}.
\newblock {\em Phys. Rev. D}, 107(8):083522, 2023.

\bibitem{Shafieloo:2016bpk}
Arman Shafieloo, Dhiraj~Kumar Hazra, Varun Sahni, and Alexei~A. Starobinsky.
\newblock {Metastable Dark Energy with Radioactive-like Decay}.
\newblock {\em Mon. Not. Roy. Astron. Soc.}, 473(2):2760--2770, 2018.

\bibitem{Pan-STARRS1:2017jku}
D.~M. Scolnic et~al.
\newblock {The Complete Light-curve Sample of Spectroscopically Confirmed SNe
  Ia from Pan-STARRS1 and Cosmological Constraints from the Combined Pantheon
  Sample}.
\newblock {\em Astrophys. J.}, 859(2):101, 2018.

\bibitem{Beutler:2011hx}
Florian Beutler, Chris Blake, Matthew Colless, D.~Heath Jones, Lister
  Staveley-Smith, Lachlan Campbell, Quentin Parker, Will Saunders, and Fred
  Watson.
\newblock {The 6dF Galaxy Survey: Baryon Acoustic Oscillations and the Local
  Hubble Constant}.
\newblock {\em Mon. Not. Roy. Astron. Soc.}, 416:3017--3032, 2011.

\bibitem{Ross:2014qpa}
Ashley~J. Ross, Lado Samushia, Cullan Howlett, Will~J. Percival, Angela Burden,
  and Marc Manera.
\newblock {The clustering of the SDSS DR7 main Galaxy sample \textendash{} I. A
  4 per cent distance measure at $z = 0.15$}.
\newblock {\em Mon. Not. Roy. Astron. Soc.}, 449(1):835--847, 2015.

\bibitem{BOSS:2016wmc}
Shadab Alam et~al.
\newblock {The clustering of galaxies in the completed SDSS-III Baryon
  Oscillation Spectroscopic Survey: cosmological analysis of the DR12 galaxy
  sample}.
\newblock {\em Mon. Not. Roy. Astron. Soc.}, 470(3):2617--2652, 2017.

\bibitem{eBOSS:2018yfg}
Gong-Bo Zhao et~al.
\newblock {The clustering of the SDSS-IV extended Baryon Oscillation
  Spectroscopic Survey DR14 quasar sample: a tomographic measurement of cosmic
  structure growth and expansion rate based on optimal redshift weights}.
\newblock {\em Mon. Not. Roy. Astron. Soc.}, 482(3):3497--3513, 2019.

\bibitem{BOSS:2017uab}
H\'elion du~Mas~des Bourboux et~al.
\newblock {Baryon acoustic oscillations from the complete SDSS-III
  Ly$\alpha$-quasar cross-correlation function at $z=2.4$}.
\newblock {\em Astron. Astrophys.}, 608:A130, 2017.

\bibitem{Chen:2018dbv}
Lu~Chen, Qing-Guo Huang, and Ke~Wang.
\newblock {Distance Priors from Planck Final Release}.
\newblock {\em JCAP}, 02:028, 2019.

\bibitem{Planck:2018vyg}
N.~Aghanim et~al.
\newblock {Planck 2018 results. VI. Cosmological parameters}.
\newblock {\em Astron. Astrophys.}, 641:A6, 2020.
\newblock [Erratum: Astron.Astrophys. 652, C4 (2021)].

\bibitem{osti_4338791}
J~Wess and B~Zumino.
\newblock {Supergauge transformations in four dimensions}.
\newblock {\em Nucl. Phys. B}, 70(1):39, 1974.

\bibitem{Brandenberger:2019jfh}
Robert Brandenberger, J\"urg Fr\"ohlich, and Ryo Namba.
\newblock {Unified Dark Matter, Dark Energy and baryogenesis via a
  \textquotedblleft{}cosmological wetting transition\textquotedblright{}}.
\newblock {\em JCAP}, 09:069, 2019.

\bibitem{Brandenberger:2019pju}
Robert Brandenberger, Rodrigo~R. Cuzinatto, J\"urg Fr\"ohlich, and Ryo Namba.
\newblock {New Scalar Field Quartessence}.
\newblock {\em JCAP}, 02:043, 2019.

\bibitem{Bertacca:2010ct}
Daniele Bertacca, Nicola Bartolo, and Sabino Matarrese.
\newblock {Unified Dark Matter Scalar Field Models}.
\newblock {\em Adv. Astron.}, 2010:904379, 2010.

\bibitem{Frion:2023}
Emmanuel Frion, David Camarena, Leonardo Giani, Tays Miranda, Daniele Bertacca,
  Valerio Marra, and Oliver Piattella.
\newblock {Bayesian analysis of Unified Dark Matter models with fast
  transition: can they alleviate the H0 tension?}
\newblock 2307.06320.

\bibitem{Callan:1977pt}
Curtis~G. Callan, Jr. and Sidney~R. Coleman.
\newblock {The Fate of the False Vacuum. 2. First Quantum Corrections}.
\newblock {\em Phys. Rev. D}, 16:1762--1768, 1977.

\bibitem{Coleman:1977py}
Sidney~R. Coleman.
\newblock {The Fate of the False Vacuum. 1. Semiclassical Theory}.
\newblock {\em Phys. Rev. D}, 15:2929--2936, 1977.
\newblock [Erratum: Phys.Rev.D 16, 1248 (1977)].

\bibitem{Marsh:2015xka}
David J.~E. Marsh.
\newblock {Axion Cosmology}.
\newblock {\em Phys. Rept.}, 643:1--79, 2016.

\bibitem{Magana:2012ph}
Juan Magana and Tonatiuh Matos.
\newblock {A brief Review of the Scalar Field Dark Matter model}.
\newblock {\em J. Phys. Conf. Ser.}, 378:012012, 2012.

\bibitem{Cicoli:2021gss}
Michele Cicoli, Veronica Guidetti, Nicole Righi, and Alexander Westphal.
\newblock {Fuzzy Dark Matter candidates from string theory}.
\newblock {\em JHEP}, 05:107, 2022.

\bibitem{Harigaya:2019qnl}
Keisuke Harigaya and Jacob~M. Leedom.
\newblock {QCD Axion Dark Matter from a Late Time Phase Transition}.
\newblock {\em JHEP}, 06:034, 2020.

\bibitem{GUTH1983321}
Alan~H. Guth and Erick~J. Weinberg.
\newblock {Could the universe have recovered from a slow first-order phase
  transition?}
\newblock {\em Nuclear Physics B}, 212(2):321, 1983.

\bibitem{PhysRevD.46.2384}
Michael~S. Turner, Erick~J. Weinberg, and Lawrence~M. Widrow.
\newblock Bubble nucleation in first-order inflation and other cosmological
  phase transitions.
\newblock {\em Phys. Rev. D}, 46:2384--2403, Sep 1992.

\bibitem{DESI:2024uvr}
A.~G. Adame et~al.
\newblock {DESI 2024 III: Baryon Acoustic Oscillations from Galaxies and
  Quasars}.
\newblock 4 2024.

\bibitem{DESI:2024kob}
K.~Lodha et~al.
\newblock {DESI 2024: Constraints on Physics-Focused Aspects of Dark Energy
  using DESI DR1 BAO Data}.
\newblock 5 2024.

\bibitem{TeppaPannia:2016hwv}
F.~A. Teppa~Pannia and S.~E. Perez~Bergliaffa.
\newblock {Evolution of Vacuum Bubbles Embeded in Inhomogeneous Spacetimes}.
\newblock {\em JCAP}, 03:026, 2017.

\bibitem{Simon:2009nb}
Dennis Simon, Julian Adamek, Aleksandar Rakic, and Jens~C. Niemeyer.
\newblock {Tunneling and propagation of vacuum bubbles on dynamical
  backgrounds}.
\newblock {\em JCAP}, 11:008, 2009.

\bibitem{Fischler:2007sz}
Willy Fischler, Sonia Paban, Marija Zanic, and Chethan Krishnan.
\newblock {Vacuum bubble in an inhomogeneous cosmology: A Toy model}.
\newblock {\em JHEP}, 05:041, 2008.

\bibitem{hawking1982bubble}
Stephen~William Hawking, IG~Moss, and JM~Stewart.
\newblock Bubble collisions in the very early universe.
\newblock {\em Physical Review D}, 26(10):2681, 1982.

\bibitem{kosowsky1992gravitational}
Arthur Kosowsky, Michael~S Turner, and Richard Watkins.
\newblock Gravitational radiation from colliding vacuum bubbles.
\newblock {\em Physical Review D}, 45(12):4514, 1992.

\bibitem{PhysRevD.84.024006}
R.~Casadio and A.~Orlandi.
\newblock Bubble dynamics: (nucleating) radiation inside dust.
\newblock {\em Phys. Rev. D}, 84:024006, Jul 2011.

\bibitem{Pannia:2021lso}
Florencia Anabella~Teppa Pannia, Santiago Esteban~Perez Bergliaffa, and Nelson
  Pinto-Neto.
\newblock {Particle production in accelerated thin bubbles}.
\newblock {\em JCAP}, 04(04):015, 2022.

\bibitem{Aguirre:2009ug}
Anthony Aguirre and Matthew~C. Johnson.
\newblock {A Status report on the observability of cosmic bubble collisions}.
\newblock {\em Rept. Prog. Phys.}, 74:074901, 2011.

\bibitem{NANOGrav:2023hvm}
Adeela Afzal et~al.
\newblock {The NANOGrav 15 yr Data Set: Search for Signals from New Physics}.
\newblock {\em Astrophys. J. Lett.}, 951(1):L11, 2023.

\bibitem{Niedermann:2019olb}
Florian Niedermann and Martin~S. Sloth.
\newblock {New early dark energy}.
\newblock {\em Phys. Rev. D}, 103(4):L041303, 2021.

\bibitem{Guth:1979bh}
Alan~H. Guth and S.~H.~H. Tye.
\newblock {Phase Transitions and Magnetic Monopole Production in the Very Early
  Universe}.
\newblock {\em Phys. Rev. Lett.}, 44:631, 1980.
\newblock [Erratum: Phys.Rev.Lett. 44, 963 (1980)].

\bibitem{Guth:1981uk}
Alan~H. Guth and Erick~J. Weinberg.
\newblock {Cosmological Consequences of a First Order Phase Transition in the
  SU(5) Grand Unified Model}.
\newblock {\em Phys. Rev. D}, 23:876, 1981.

\bibitem{Guth:1980zm}
Alan~H. Guth.
\newblock {The Inflationary Universe: A Possible Solution to the Horizon and
  Flatness Problems}.
\newblock {\em Phys. Rev. D}, 23:347--356, 1981.

\bibitem{Coleman:1980aw}
Sidney~R. Coleman and Frank De~Luccia.
\newblock {Gravitational Effects on and of Vacuum Decay}.
\newblock {\em Phys. Rev. D}, 21:3305, 1980.

\end{thebibliography}

\end{document}